\documentclass{article}
\usepackage[a4paper, margin=0.71in]{geometry}
\usepackage{multicol}
\usepackage{amsmath}
\usepackage{graphicx}
\usepackage{caption}
\usepackage{subcaption}
\usepackage{float}
\usepackage{enumitem}
\usepackage[font=small,labelfont=bf]{caption}
\title{\textbf{Upconversion of infrared light by graphitic micro-particles due to photo-induced structural modification}}
\author{Rohin Sharma\textsuperscript{1,$\dagger$}, 
Nishma Bhattarai\textsuperscript{1}, 
Rijan Maharjan\textsuperscript{1}, 
Lilia M. Woods\textsuperscript{2}, 
Nirajan Ojha\textsuperscript{1},\\
Ashim Dhakal\textsuperscript{1,*}}
\date{}
\begin{document}
\maketitle
\noindent
\emph{\textsuperscript{1} Biophotonics Lab, Phutung Research Institute, PO Box 12335, Kathmandu 44600, Nepal.}\\
\emph{\textsuperscript{2} University of South Florida, Department of Physics, Tampa, FL 33620, USA.}\\
\noindent
\\
$\dagger$ Current email: rohin@usf.edu\\
* Corresponding email: ashim.dhakal@pinstitute.org
\paragraph{}
\paragraph{}
\textbf{Abstract}: Recent reports of upconversion and white light emission from graphitic particles warrant an explanation of the physics behind the process. We offer a model, wherein the upconversion is facilitated by photo-induced electronic structure modification allowing for multi-photon processes. As per the prediction of the model, we experimentally show that graphite upconverts infrared light centered around 1.31~$\mu$m to broadband white light centered around 0.85 $\mu$m. Our results suggest that upconversion from shortwave infrared ($\sim$3~$\mu$m) to visible region may be possible. Our experiments show that the population dynamics of the electronic states involved in this upconversion process occur
in the timescale of milliseconds.
\vspace{5mm}
\hrule
\begin{multicols}{2}
\section*{Introduction}
Graphene has been identified as a material with remarkable optical and electronic properties suitable for various opto-electronic applications such as photocurrent measurement and sensing ~\cite{xia2009photocurrent, doi:10.1021/ph400147y, MAK20121341, gabor2011hot}. Consequently, optical and electronic properties of graphitic materials have been the topic of great interest. As a zero band gap material, graphite and pristine graphene is expected to have poor emission properties. However, several experiments have reported broad-band radiation in the visible and near-infrared region from graphite and graphene as a result of ultrafast as well as continuous wave photonic excitation~\cite{PhysRevB.82.081408, PhysRevLett.105.127404, PhysRevLett.108.167401, strek2015laser, 10.1063/1.2980321, strek2017laser}. To explain this phenomenon, a number of studies investigated the carrier lifetime dynamics in graphite and graphene through THz spectroscopy, pump-probe spectroscopy and theoretical methods~\cite{PhysRevLett.102.086809, PhysRevB.77.121402, PhysRevLett.87.267402, PhysRevB.42.2842, shibuta2016direct, 10.1063/1.2956669, PhysRevB.76.115434, PhysRevLett.101.077401, ishida2011non, PhysRevLett.111.027403, tielrooij2013photoexcitation, PhysRevB.92.165429, PhysRevB.87.045427, PhysRevB.60.17037}. Significant experimental studies focused on the image potential states~\cite{PhysRevB.87.045427, PhysRevB.60.17037, PhysRevB.89.155303, Gugel_2015}, but the true mechanism of the carrier excitation remains controversial as appropriate initial states for vertical excitation (direct band-gap transition) does not exist. Several features such as lower sample temperature than that predicted by the theory of black-body radiation \cite{10.1063/1.2980321, PhysRevX.7.011004}, frequency-independent absorption and non-linear dependence of emission intensity with the excitation power scaled effectively by the power law $P^N$ (N being order of the process) serves as a direct evidence towards the multi-photon photoemission process~\cite{strek2015laser, strek2017laser, PhysRevX.7.011004}.

Investigation of carrier generation and thermalization in graphite through ultrafast time resolved multiphoton spectroscopy confirms that the redistribution of the $\pi$ electrons with bonding characters to the band extrema; namely the $\sigma$-symmetry inter-layer band ($\sigma$-ILB) at the $\Gamma$ point through $e$-phonon scattering causes the lattice instability to form distorted structure as part of the multi-photon photoemission process~\cite{PhysRevLett.87.267402, PhysRevX.7.011004}. But there lacks a study explicitly showing the $\sigma-\pi$ mixing or coupling of the $\pi$ and $\sigma$ bands in the electronic structure, as electron transfer between the $\pi$ and $\sigma$ band cannot occur if the bands are completely decoupled as in the case of pure graphite and graphene. Since the optical properties of graphite are entirely described by the $\pi\pi^*$ electrons, investigating the electronic band structure of structurally distorted graphitic phase and deducing the optical properties may enable us to gain deeper insights into the laser--induced emission mechanism in graphite.

Moreover, several experiments have pointed to an existence of optically induced intermediate, metastable graphitic phases exhibiting local $sp^3$ bonded tetrahedron network typically stable at high pressure environment~\cite{PhysRevB.77.121402, PhysRevLett.101.077401, narayan2015novel, PhysRevB.80.014112}. Furthermore, electron crystallography studies investigating into the dynamics of induced structural changes have reported interlayer compression and in-plane shear displacement between the graphene layers~\cite{PhysRevLett.101.077401, PhysRevLett.100.035501} as a result of increased population of $C2p_z$ orbitals hybridizing to $pp\sigma$ bonding states, resulting in connection of the neighbouring layers. Further corroboration comes from scanning tunnelling microscopy study and \emph{ab-initio} study suggesting $p$ polarized light playing a role in enhancement of such local distortions via interlayer charge transfer excitation induced by the electric vector component~\cite{PhysRevLett.102.087402, Nakayama_2002}.

In this study, we have performed first-principle calculation using density functional theory to simulate the electronic band structure and studied the optical properties of graphite and photo-induced, $sp^2$--$sp^3$ hybridized graphite. This paper serves as a computational work confirming the speculations of photo-induced phase transition from $sp^2$ to $sp^2$--$sp^3$ matastable state to explain the origin of the multiphoton-induced white light emission~\cite{strek2015laser, strek2017laser}. The predicted optical emission of the distorted structure comply well with our experimentally observed broad-band emission spectra from graphite microparticles under excitation from a swept source pulsating infrared laser centered at 0.95 eV energy. This is also a first report of the upconversion phenomenon in graphite from an O-band infrared radiation source. Finally, we have analysed the emission and decay nature of the broad band emission under modulating signal of CW Near-Infrared excitation of wavelength 785 nm, the results of which can give some insights on the population dynamics of the carriers during the photoemission process.

\end{multicols}
\begin{figure} [H]
     \centering
     \begin{subfigure}[b]{1\textwidth} 
     \centering
         \includegraphics[width=14cm]{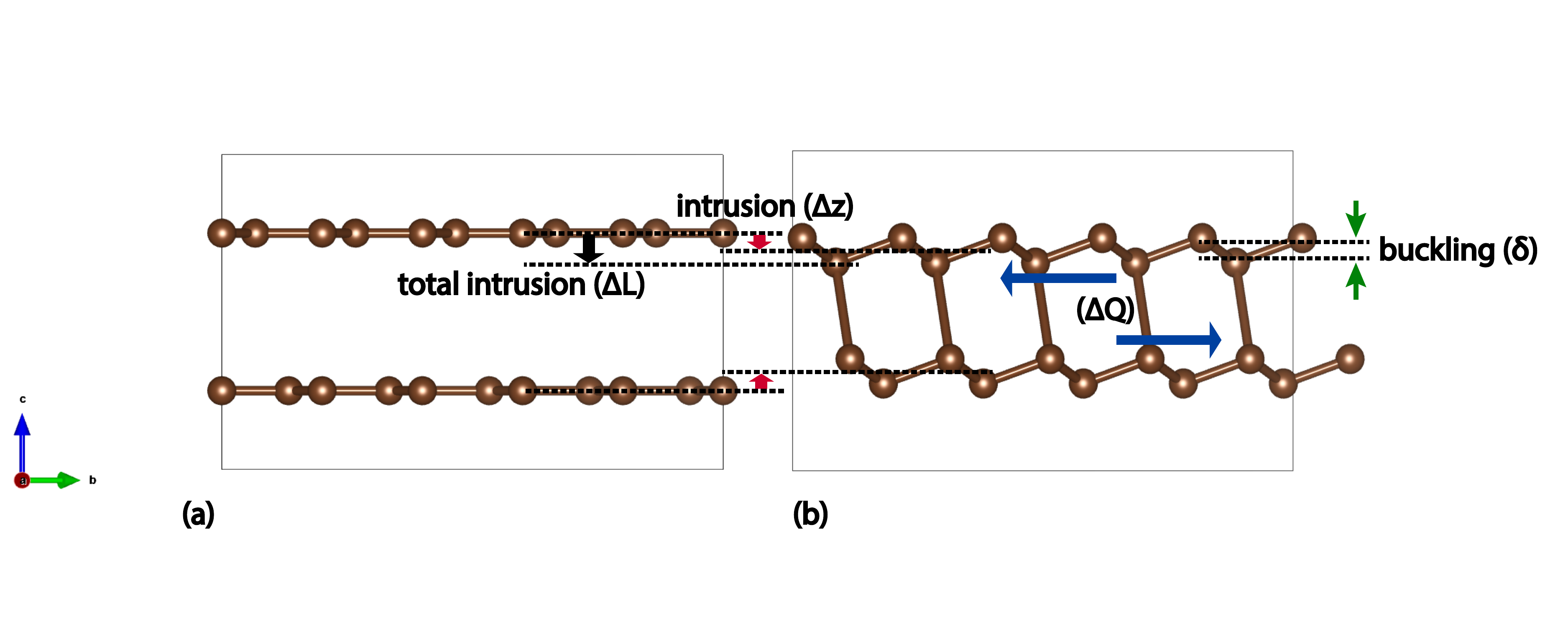}
     \end{subfigure}
     \hfill
     \begin{subfigure}[b]{1\textwidth}
     \centering
         \includegraphics[width=14cm]{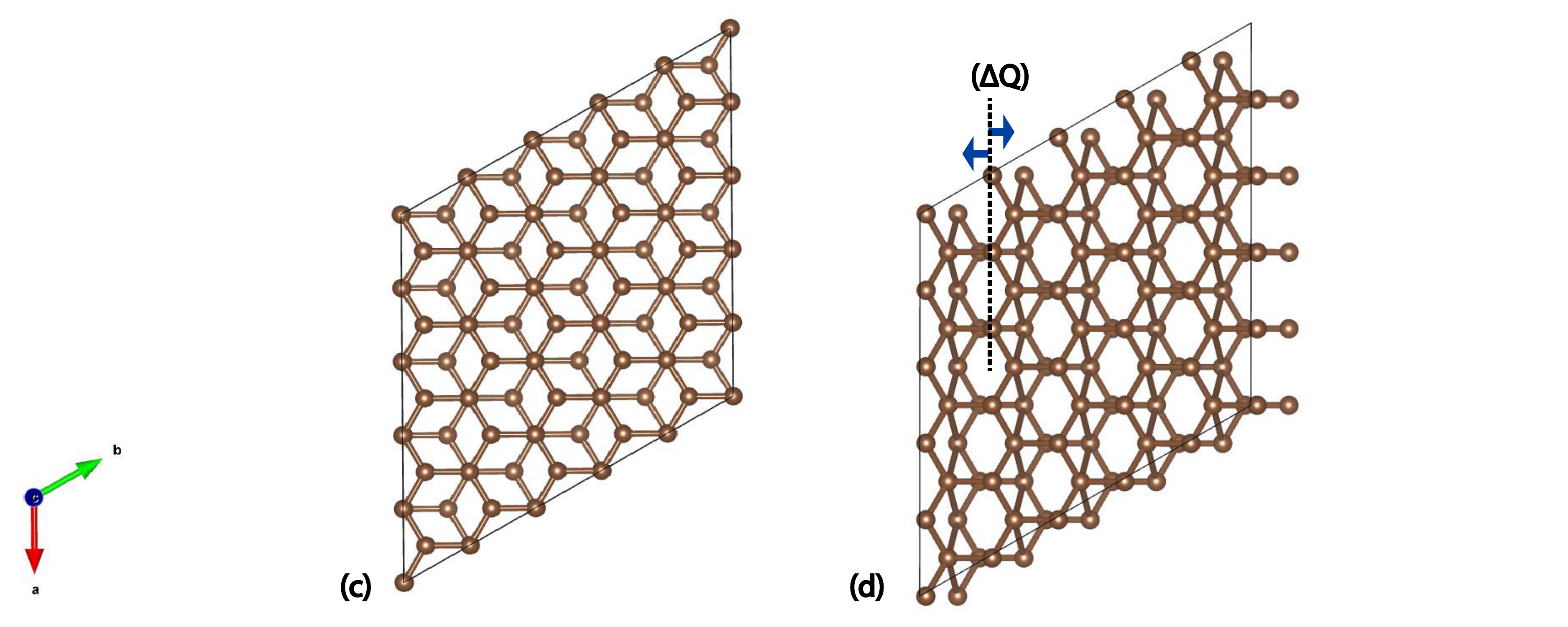}
         \label{BuckUnit}
     \end{subfigure}
     \caption{Atomic position visualisation generated from VESTA (Visualisation for Electronic Structural Analysis). (a)~Graphite supercell in 
     $sp^2$ hybridization with an interlayer spacing of 3.34~$\AA$. (b)~Photo-induced structural deformation of graphite showing shear ($\Delta Q$), intrusion ($\Delta z$) and buckling ($\delta$) of graphene layers resulting in total intrusion ($\Delta L$) between the layers producing local $sp^3$ hybridization. (c)~Graphite supercell showing $AB$ stacking along $c$ axis. (d)~Deformed graphite supercell viewed along $c$ axis with the optimised distortion parameters $\Delta Q$ and $\Delta L$.}
\label{GraphiteUnit}
\end{figure}
\begin{multicols}{2}
\section*{Numerical analysis}
Our calculations relies on VASP (Vienna \emph{Ab-initio} Simulation Package) codes~\cite{PhysRevB.47.558} based on Density Functional Theory formalism where the valence electrons are treated in Local Density Approximation. The approach is based on an iterative solution of the Kohn-Sham equation~\cite{PhysRev.140.A1133} in a plane-wave basis set with the Projector-Augmented Wave pseudopotential~\cite{PhysRevB.59.1758, PhysRevB.50.17953}. Our system consists of $AB$ stacked bi-layer graphene separated by a distance of 3.34 $\AA$ with 2 atoms per layer in the unit cell (Fig. \ref{GraphiteUnit}). The total energy calculation was done with a plane wave basis set in a dense k-point mesh of  $15 \times 15 \times 6$ and with the plane wave kinetic energy cutoff of 450 eV. A full structural relaxation was done initially to establish the ground state through energy minimization. Then the electronic structure calculations were performed self-consistently to produce the energy band structure of graphite (Fig. \ref{graphite}). 

Upon laser irradiation, the increased population in the $2Cp_z$ orbital and depopulation of the in-layer bonding states results in interlayer attraction resulting in intrusion $\Delta z$, and inlayer expansion resulting in shear $\Delta Q$~\cite{PhysRevLett.101.077401}. Additionally, the deviation from the $sp^2$ to $sp^2$--$sp^3$ hybrid configuration requires buckling $\delta$, where the atoms are alternatively displaced out of plane in each layer~\cite{PhysRevLett.102.087402, radosinski2012photoinduced}. So three modes of displacements are necessary to model the out-of-equilibrium graphitic structure. Starting with a buckling of 0.5~$\AA$, we gradually introduced the intrusion such that the total intrusion $\Delta L = \Delta z + \delta/2$ increased from 0.6 to 0.7~$\AA$ with a 0.005 $\AA$ increment. Then the shear $\Delta Q$ was introduced in each intrusion upto 0.4~$\AA$ with an increment of $0.05 \AA$. Optimization of the total intrusion $\Delta L$ and shear $\Delta Q$ was done through energy minimization. Implementing the optimized $\Delta L$ and $\Delta Q$, the electronic structure calculation was performed to obtain the energy-band diagram (Fig. \ref{buck}). The optical properties of pure and distorted graphite were calculated using a dense k-point grid within the Random Phase Approximation approach in which the local field effects are included at Hartree level only. The frequency dependent dielectric matrix is calculated after the electronic ground state has been determined, which can be written as the sum of real and imaginary part: $\epsilon = \epsilon^{(1)} + i\epsilon^{(2)}$, the imaginary part of which is determined by a summation over empty states using the equation:
\begin{multline}
    \varepsilon_{\alpha\beta}^{(2)}(\omega) = \frac{4\pi^2e^2}{\Omega}lim_{q\rightarrow 0}\frac{1}{q^2}2\omega_k\delta(\epsilon_{ck} - \epsilon_{v k} - \omega) \\ 
    \langle u_{ck+e_{\alpha} q} | u_{v k}\rangle \langle u_{ck + e_{\beta q}} | u_{v k} \rangle,
\end{multline}
where the indices $\alpha$ and $\beta$ are the cartesian components, vectors $e_{\alpha}$ and $e_{\beta}$ are the unit vectors along three direction, $c$ and $v$ refers to conduction and valence band and $u_{ck}$ is the cell periodic part of the orbitals at the k-point. The imaginary part of the dielectric function for in plane polarization, $\epsilon_{xx}^{(2)}$ obtained in this way, which directly gives the absorption spectra of $sp^2$--$sp^3$ hybridized graphite is presented in 
Fig.~\ref{graphite_emission}. As the real part describes the materials influence on the propagation of light, the ratio $\epsilon^{(2)}/\epsilon^{(1)}$ can provide some information about the energy dissipation and dielectric properties. However, since this study is mainly concerned with the light absorption and emission properties, the real part is not of prime focus in this work.

\end{multicols}
\begin{figure} [H]
     \centering
     \begin{subfigure}[b]{0.45\textwidth}
         \centering
         \includegraphics[width=\textwidth]{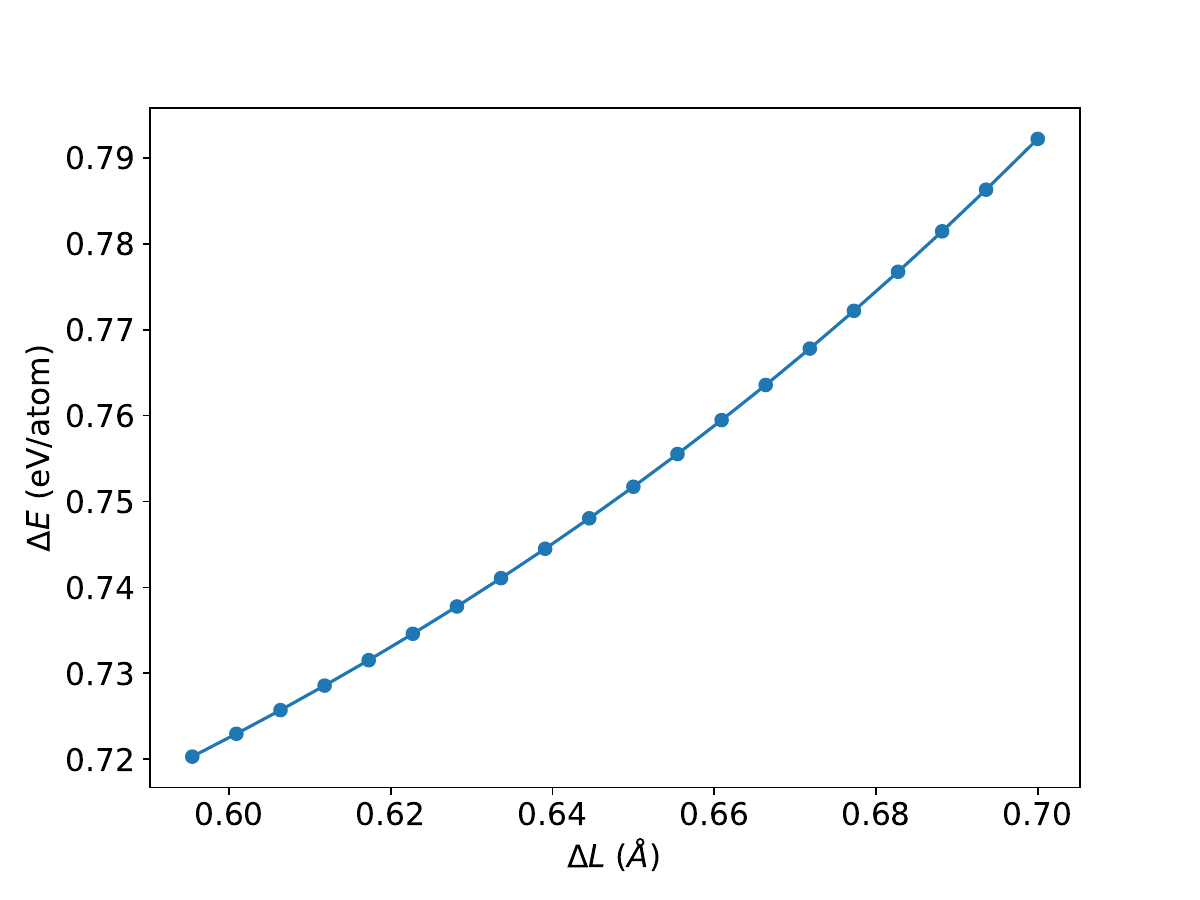}
         \caption{}
         \label{delL}
     \end{subfigure}
     \hfill
     \begin{subfigure}[b]{0.45\textwidth}
         \centering
         \includegraphics[width=\textwidth]{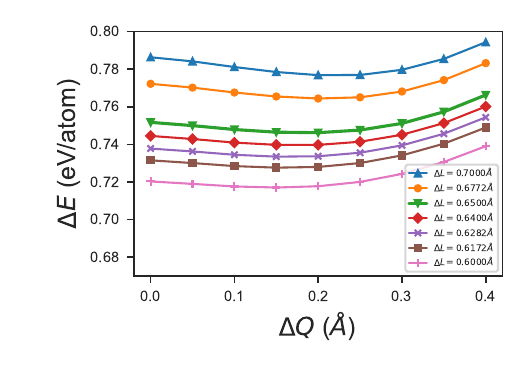}
         \caption{}
         \label{delQ}
     \end{subfigure}
\caption{Structural optimization through energy minimization process. (a)~The difference in total energy from that of graphite is plotted as a function of total intrusion $\Delta L$. (b)~Through optimization of the shear, $\Delta Q = 0.2$~$\AA$ at $\Delta l = 0.64$~$\AA$ is taken as the optimum distortion parameters for the $sp^2$--$sp^3$ hybridized graphite.}
\label{Optimize}
\end{figure}
\begin{figure} [H]
     \centering
     \begin{subfigure}[b]{0.47\textwidth}
         \centering
         \includegraphics[width=0.88\textwidth]{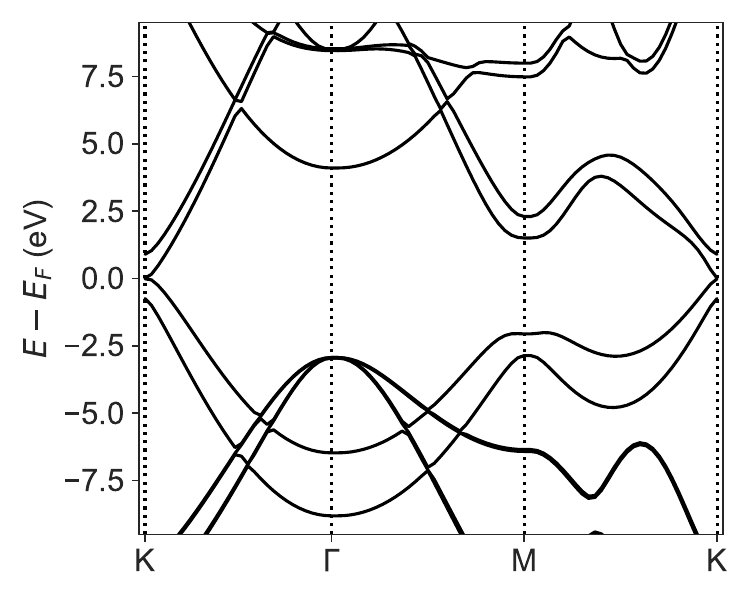}
         \caption{}
         \label{graphite}
     \end{subfigure}
     \hfill
     \begin{subfigure}[b]{0.47\textwidth}
         \centering
         \includegraphics[width=0.88\textwidth]{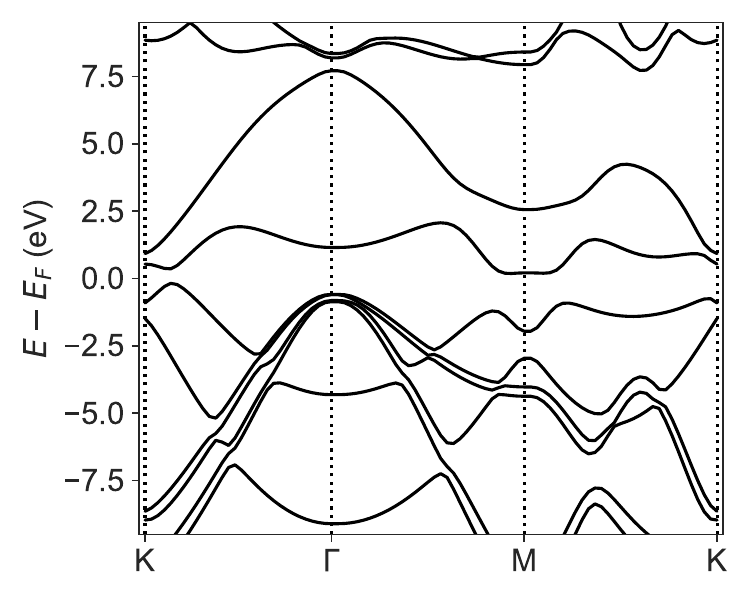}
         \caption{}
         \label{buck}
     \end{subfigure}
\caption{Electronic band structure of graphite and structurally distorted graphite. (a)~Electronic band structure of graphite showing $\pi$ and $\sigma$ bands. The $\pi$ and $\sigma$ band are decoupled indicating no $\sigma$--$\pi$ mixing. The band gap at $k$ point is zero. (b)~The electronic structure of structurally deformed graphite showing $\sigma$--$\pi$ mixing. A direct band gap of 0.54~eV is seen near the k-point.}
\label{electronic}
\end{figure}
\begin{figure} [H]
     \centering
     \begin{subfigure}[b]{0.47\textwidth}
         \centering
         \includegraphics[width=9cm]{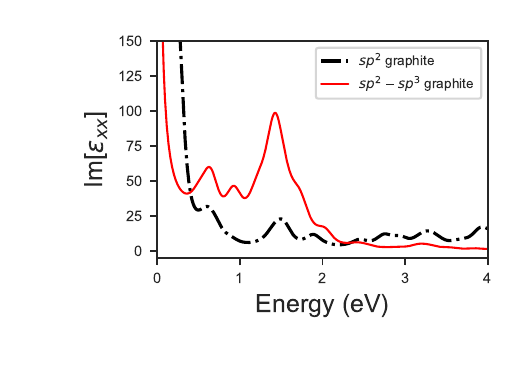}
         \caption{}
         \label{graphite_emission}
     \end{subfigure}
     \hfill
     \begin{subfigure}[b]{0.47\textwidth}
         \centering
         \includegraphics[width=9cm]{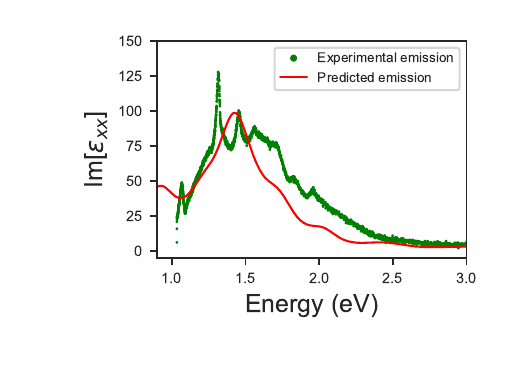}
         \caption{}
         \label{emission}
     \end{subfigure}
\caption{Optical properties of pure and structurally distorted graphite calculated using density functional theory. (a)~Calculated absorption spectra of $sp^2$ structure and $sp^2$--$sp^3$ hybrid structure with parameters $\Delta L = 0.64\AA$, $\Delta Q = 0.2\AA$ and $\delta = 0.5\AA$. The major absorption peak in the visible-IR range is at 1.43~eV and minor peaks are at 0.92~eV and 0.62~eV. (b)~Experimental anti-stokes emission compared with the calculated emission of the $sp^2$--$sp^3$ hybrid configuration. The sharp peak at 1.31~eV is due to the OH-absorption of the optical fiber used to collect the signal.}
\label{opticalproperties}
\end{figure}
\begin{multicols}{2}
\section*{Experimental results}
We analysed the broadband anti-stokes emission in the visible and NIR range from a highly oriented pyrolytic graphite poweder of size $<$ 45~$\mu$m (Nanographi Nanotech, UK) with the excitation in the Infrared range. To study the emission nature, it was necessary to sustain the emission by preventing oxidation. To achieve this the graphite powder mixed with 1~ml of deionized water was dropcasted in a microscopic slide. After evaporation of water (heating in hotplate 100 \textdegree C for 5 mins), it was covered by a $\sim 25$~$\mu$m thin  polydimethylsiloxane layer (spin-coated and heat-cured) to inhibit oxidation. Following two cases were studied.\\

\begin{figure} [H]
     \centering
     \begin{subfigure}[b]{0.48\textwidth}
         \centering
         \includegraphics[width=7cm]{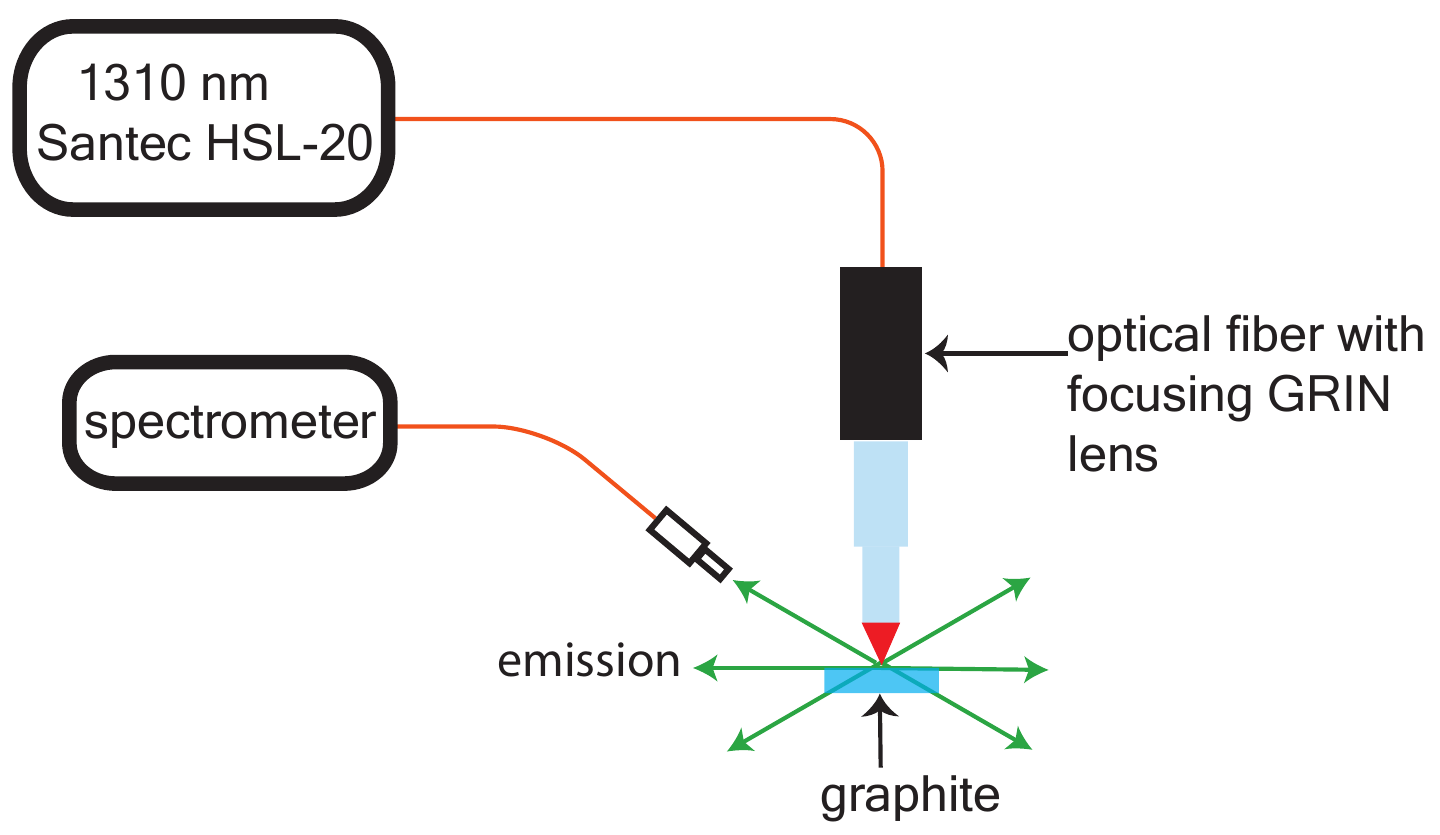}
         \caption{}
         \label{schematic1}
     \end{subfigure}
     \hfill
     \begin{subfigure}[b]{0.5\textwidth}
         \centering
         \includegraphics[width=5cm]{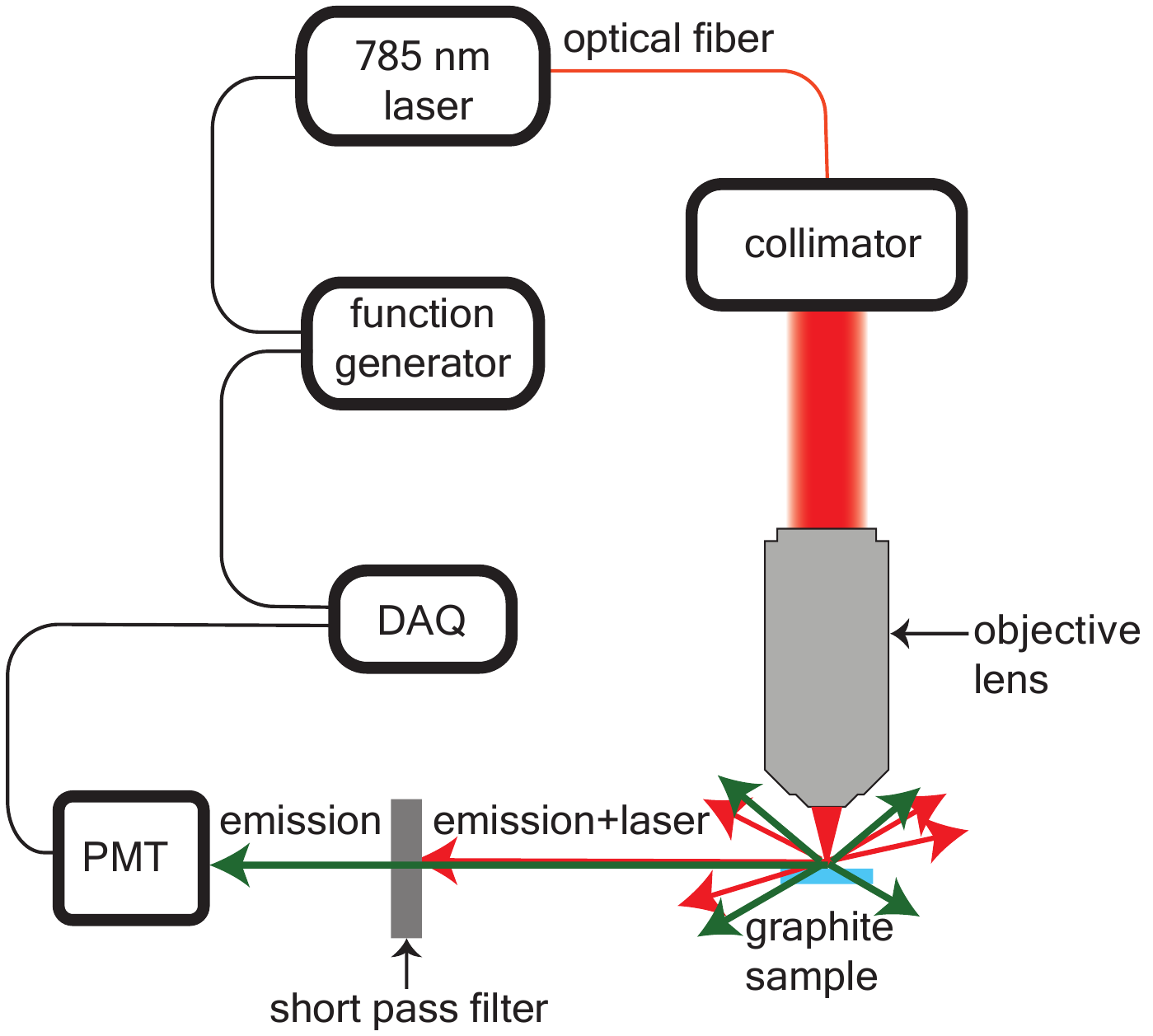}
         \caption{}
         \label{schematic2}
     \end{subfigure}
\caption{Schematics of the experimental setups: (a)~for detection of WLE from the Santec HSL-20 infrared laser source,(b)~to study the temporal dynamics of emission and decay from a modulating NIR laser source.}
\label{emission_decay_nature}
\end{figure}

Case 1: To study the spectral properties of WLE from a O-band excitation, we used 1310 nm excitation spectrum (Santec HSL-20, 20mW average incident power) as the source as shown in Fig.~\ref{schematic1}. The emission spectrum was collected by an optical fiber positioned at an angle with the sample which was then acquired with ASEQ spectrometer (LR1-B).\\

Case 2: To study the temporal nature of excitation and decay of the emission, we modulated a 785 nm laser (CNI-Laser FC-D-785, 420 mW ) as an NIR excitation source to achieve an excitation duration of  40~ms. Focusing the excitation laser beam through an objective lens, the anti-stokes emission was detected by a photo-multiplier tube from a direction perpendicular to the incident laser as shown in the Fig.~\ref{schematic2}.

As predicted by our numerical analysis, we observed an emission in Case 1, with the peak intensity in the visible region accompanied by absorption of 1310~nm infrared radiation~(Fig.~\ref{emission}). This up-conversion process has been made possible due to the change in optical properties of graphite brought about by the structural modification from $sp^2$ graphite to $sp^2$--$sp^3$ hybridized graphite upon photo-excitation.
\end{multicols}

\begin{figure} [H]
     \centering
     \begin{subfigure}[b]{0.47\textwidth}
         \centering
         \includegraphics[width=8.5cm]{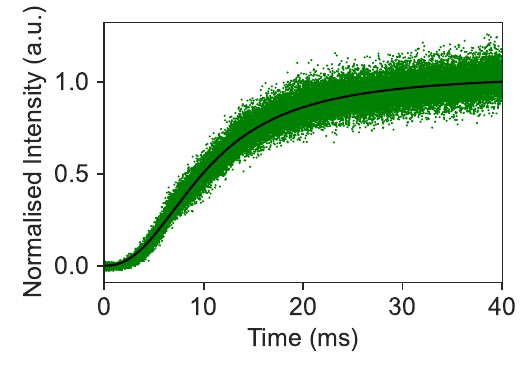}
         \caption{}
         \label{decay}
     \end{subfigure}
     \hfill
     \begin{subfigure}[b]{0.47\textwidth}
         \centering
         \includegraphics[width=8.5cm]{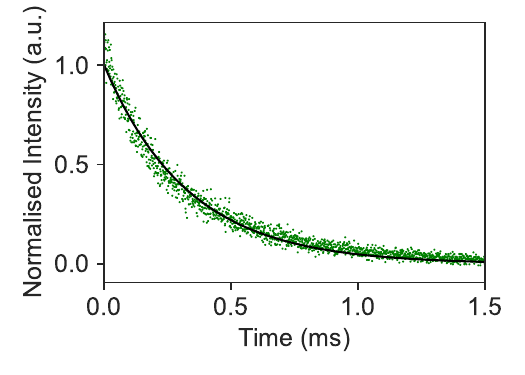}
         \caption{}
         \label{expo_decay}
     \end{subfigure}
\caption{(a)~Visible-NIR anti-stokes emission growth curves observed from graphite excited with 1.58 eV (785 nm) CW laser pulse of 40 ms duration and 420 mW power. (b)~Life time of the anti-stokes emission when the laser in the OFF-state showing exponential decay with a time constant of 0.33 ms.}
\label{emission_decay_nature}
\end{figure}
\begin{multicols}{2}
\section*{Results and discussion}
From the total energy calculations using DFT calculations, we have optimized the shear and intrusion of the graphene layers to model the distortion brought about by the laser incidence onto the graphite sample. It is seen in Fig.~\ref{delQ} that for a total intrusion $\Delta L \ge 0.64$~$\AA$, stability is seen at the shear of 0.2~$\AA$ while for lower intrusions, the stability is seen at the shear of 0.15~$\AA$. Since these parameters are close to the Scanning Tunneling Microscopy measurement results~\cite{PhysRevLett.102.087402} and large scale Molecular Dynamics simulation results~\cite{radosinski2012photoinduced}, we expect these to be the distortion parameters of photo-induced structurally deformed, $sp^2$--$sp^3$ hybridized graphite. Fig.~\ref{graphite_emission} shows the expected optical emission (which would be the same as absorption spectra under resonance condition) of the distorted graphite, showing a broad band emission with a peak intensity at 1.43~eV. This is in good agreement with our experimentally observed broad-band anti-stokes emission with a peak intensity at 1.45~eV upon excitation with an infrared source. The minor absorption peaks in the infrared region indicates that upconversion can be expected form as low as 3~$\mu$m excitation wavelength, albeit less efficiently. In addition to the peak overlap, a close correspondence of the breadth and minor peaks between the predicted optical emission and the experimentally observed emission peak in Fig.~\ref{emission} adds more corroboration to the claim that the observed emission is indeed an emission from distorted $sp^2$--$sp^3$ hybridized graphite. The sharp peak in the experimental data at 1.31~eV seen in Fig.~\ref{emission} is from the material of the fiber (OH-absorption) that is used in the setup to receive the signal. 

Hence, our results  suggests that upon incidence of an infrared laser, there is a formation of local $sp^2$--$sp^3$ hybridized domain with the distortion parameters of $\Delta L = 0.64$~$\AA$ and $\Delta Q = 0.2$~$\AA$, such that the optical property is altered to emit a prominent broad band anti-stokes emission in the visible and NIR region. This also demonstrates that white light emission from graphite is not only possible from ultrafast--NIR lasers but also with excitation from an infrared and NIR CW lasers~\cite{strek2015laser, 10.1063/1.2980321, strek2017laser}. 

This alteration of the optical properties can clearly be interpreted by analysing the changes in the electronic band structure. The transition from $sp^2$ hybridized graphite to a $sp^2$--$sp^3$ hybrid structure results in significant modifications of the electronic structure that alters the optical properties of the material. As seen in Fig.~\ref{buck}, first there is an opening of a band gap of 0.54~eV near the k-point making it possible for the absorption of a photon or multiple photons  through the multiphoton process to facilitate electronic excitation. Then the $\pi$ and $\sigma$ bands are coupled indicating $\sigma$--$\pi$ mixing and making electron transfer from $\pi$ to $\sigma$ bands possible. This provides a route for the relaxation of the $\pi$ electrons to the band extrema as indicated in Ref.~\cite{PhysRevX.7.011004} through e-phonon scattering. Finally the $\pi^*$ band is lowered in energy such that the band gap at the $\Gamma$ point reaches around 1.7~eV that makes the emission of photon of energy in the visible and NIR wavelength range possible.

The nature of the anti-stokes emission growth observed in Fig 6, can be modeled  with a logistic-growth equation 
\begin{equation} \label{logisticeq}
    I = I_o \left[1 - \frac{1}{1 + \left(\frac{t}{\tau_a}\right)^b}\right],
\end{equation}
where $I_o$ is the maximum intensity, $\tau_a$ is the inflection points where intensity falls to half and $b$ is the steepness of the curve. The experimental decay curve of the emission can be approximated by an exponential function of the form. 
\begin{equation} \label{decayeq}
    I = I_o e^{-{\frac{t}{\tau_d}}}.
\end{equation}

The values of $\tau_a$, $\tau_d$ and $b$ are experimentally determined to be  10 ms, 0.33 ms and 2.33 ms respectively. These results show that the population dynamics of the electronic states during photo-induced phase transitions occur in the scale of milliseconds. 

The basic mechanism leading to the observed emission dynamics can be understood through a non-adiabatic transition model where population in the energy-levels are given by the Landau-Zener diabatic transition rate~\cite{landau1932theory, zener1932non}. The energy of the photo-induced electrons in the excited state gets dissipated to the environment non-adiabatically. The time evolution of the dissipation rate can be obtained by constructing the time-evolving density matrix of the system which can be used to extract the time-dependent population probability~\cite{zanca2018frictional, arceci2017dissipative}. The time dependent population probability follows exponential decay from excited state to the ground state during spontaneous emission process resulting in an exponentially decaying emission intensity corresponding to Fig.~\ref{expo_decay} and shows a logarithmic growth from excited state to the ground state during the stimulated emission process resulting in the logistic emission nature corresponding to Fig.~\ref{decay}.

These results altogether will help future experimental and theoretical studies regarding the kinetics of the charge carriers during the photo-excitation process. 

\section*{Conclusion}
We experimentally demonstrate up-conversion in graphite from an O-band infrared excitation to NIR band and theoretically explain the results. Our  model suggests that photo-induced phase transition of graphite into its metastable state where the graphene layers are distorted to form local $sp^3$--hybridized structures. This transformation (a) opens the band gap near the k-point, (b) lowers the band gap at the $\Gamma$ point, and (c) allows $\sigma$--$\pi$ mixing, thereby creating the optically--induced out--of--equilibrium graphitic phase which has significant absorption in the infrared region and emission in the visible-NIR region. Our results suggests the possibility of the up-conversion process in graphite from an infrared source from down to 3 $\mu$m wavelength range to visible-NIR range. These results may open new applications utilizing the up-conversion process from infrared lasers, which are ubiquitous, thanks to their use in telecommunication and sensing applications.
\section*{Acknowledgement}
This work was supported by Grants: The World Academy of Sciences (18-013RG/Phys/AS\_I, 21-334 RG/PHYS/AS\_G, 22-244RG/PHYS/AS\_G). NB and AD likes to acknowledge V. K. Jha and B. Adhikari at St. Xaviers College, Kathmandu, Nepal for some discussions.
\vspace{3mm}
\hrule
\bibliography{ref}
\bibliographystyle{ieeetr}
\end{multicols}
\end{document}